\documentclass[aps,prc,twocolumn,showpacs,preprintnumbers,amsmath,amssymb]{revtex4}
\usepackage{graphicx}
\usepackage{graphicx}% Include figure files
\usepackage{dcolumn}% Align table columns on decimal point
\usepackage{bm}% bold math

\def\Rapt{{R_{AA}(p_T)}}
\def\Ra{{R_{AA}}}
\def\v2{{v_{2}(p_T)}}

\def\vb{{\vec b}}
\def\vr{{\vec r}}

\begin{document}

\title{Sensitivity of the Jet Quenching Observables to\\ the Temperature Dependence of the Energy Loss}

\author{Scardina Francesco$^{1,2}$}\email{scardinaf@lns.infn.it}
\author{Massimo Di Toro$^{2,3}$}
\author{Vincenzo Greco$^{2,3}$}\email{greco@lns.infn.it}
\affiliation{$^1$Department of Physics, University of Messina, I-98166 - Messina.\\
$^2$Department of Physics and Astronomy, University of Catania, Via S. Sofia 64, I-95125 Catania.\\
$^3$INFN- Laboratori Nazionali del Sud, Via S. Sofia 62, I-95125 Catania}

\date{\today}

\begin{abstract}
The quenching of minijet (particles with $p_T>> T, \Lambda_{QCD}$ ) in ultra-relativistic heavy-ion collisions
has been one of the main prediction and discovery at RHIC. We analyze the correlation between
different observables like the nuclear modification factor $\Rapt$, the elliptic flow and the ratio of
quark to gluon suppressions. We show that the temperature (or entropy density) dependence 
of the in-medium energy loss strongly affects the relation among these observables. In particular
the large elliptic flow and the nearly equal $\Rapt$ of quarks and gluons can be accounted for only if
the energy loss occurs mainly around $T_c$ and the $q\leftrightarrow g$ conversion is significant. 
The use of an equation of state fitted to lattice QCD calculations, slowing down the cooling
as $T \rightarrow T_c$, seems to contribute to both the enhancement of $v_2$ and the efficiency of the
conversion
mechanism.

\end{abstract}


\maketitle

%\section{Introduction} 

The experiments at the Relativistic Heavy Ion Collider (RHIC) are dedicated to study the properties of the
matter at exceedingly high density and temperature. At such extreme conditions the matter was expected to
undergo a deconfinement and chiral phase transition and have quark and gluons as degrees of freedom.
The theoretical and experimental efforts have shown that indeed a new form of matter has been created 
\cite{rhic_white_paper}.
Such a matter appears as a nearly perfect fluid with very low viscosity to entropy density 
\cite{Romatschke:2007mq,Heinz:2009cv}. It develops strong
collective modes with quarks as degrees of freedom and hadronizes in a modified way
respect to pp collisions, at least in the intermediate $ 2 \leq p_T \leq 5$ GeV region
\cite{Fries:2008hs,Greco:2007nu}. One way to probe the created matter is to exploit the high energy jets
($p_T >> T, \Lambda_{QCD}$) produced by the hard collisions at the initial stage. They are internal probes
propagating through the
fireball and interacting  with the medium, hence carrying information on its properties as proposed long
ago in Ref.s \cite{Bjorken:1982tu,Appel_Blaizot_McLerran,Gyulassy:1990ye}. It has indeed been shown that the
matter has a very
high opacity respect to high $p_T$ partons that traverse the hot medium in agreement with the expectations about the energy loss in QCD medium \cite{Gyulassy:1990ye,Baier:1996sk,Gyulassy:2003mc}. 
This energy loss can be quantified by the suppression of observed hadron
spectra at high transverse momenta $p_T$, as well as in the suppression of back-to-back di-hadron correlations
with a high-$p_t$ trigger, when compared with pp or d-A collisions \cite{rhic_white_paper,Phenix_more}. 
Both these phenomena related to the ``jet
quenching'' have been observed and represent one of the major discoveries of the RHIC experimental
program \cite{rhic_white_paper}. 

However even if the observation of the jet suppression cannot be questionated
there are several fundamental questions that still remain open. 
There are indeed several models that depend in a different way on temperature 
$T$ \cite{dglv,amy}, or that do not depend explicitly on $T$ but on the $\hat q$
transport coefficient\cite{bdmps,asw}, some others are based on perturbative approach 
other on higher twist expansion \cite{higher_twist1,higher_twist2}.

Despite differences all the approaches seems to be able to describe the amount of suppression $\Rapt$
observed experimentally.
It is patent that one needs to go one or two steps further and fortunately experimentally there is already the
availability of other observables related to the jet quenching phenomena. Interestingly the present models does
not seem to be able to account for all of them simultaneously. 
In this paper we will focus on two observables beyond the
$\Rapt$. One is the elliptic flow $v_2(p_T)$ and the other is the flavor dependence of the suppression that we
will mainly discuss in terms of $\Ra(q)/\Ra(g)$. The latter can be experimentally 
inferred by a systematic comparison of the different
suppression for $\pi,\rho,K,p,\bar p$ that are differently related to quark and gluon
suppression.

The purpose of the present paper is to show that the jet-quenching mechanism carries much more information
than what can be inferred from only the nuclear modification factor $\Rapt$
and even the jet-triggered angular information. 
We suggest that the study of the correlation between the elliptic flow $\v2$ and the flavor dependence of the
quenching $R_{AA}(q)/R_{AA}(g)$ is rich of information on the temperature dependence of the quenching and 
on the mechanism of parton flavor conversion. 

%A first comparison with experimental results show that the present
%model of jet quenching can hardly describe the observations. However 
We point out that a correlation
between $\v2$ and $R_{AA}(q)/R_{AA}(g)$ is sensitive to the temperature (or entropy density $s$)
dependence 
of $\Delta E_{loss}$ and on the density profile of the bulk.
The latter becomes unexpectedly dramatic if an extreme $T$, $\rho$ or $s$ dependence is
considered as in the recent works of E. Shuryak and J. Liao \cite{LS_PRL} or of V.S. Pantuev
\cite{Pantuev:2005jt}. 
This shows in general that once one goes beyond the
$\Rapt$ a more careful treatment of the time evolution of the fireball becomes mandatory, but also the
simultaneous description of $\Rapt$, $\v2$, and $R_{AA}(q)/R_{AA}(g)$ 
(or the ratio of the $\Rapt$ between different hadrons) is non trivial and more rich of information.
In particular we find that an energy loss that increase as $T\rightarrow T_c$,
the $q \leftrightarrow g$ in-medium conversion and the expansion-cooling of the fireball according to a
lattice QCD EoS all go in the direction of improving the agreement with the three observables.

The paper is organized in five Sections. In Section I
the main ingredients of the model are presented. In Section II the model is applied to calculate the $\Rapt$ as a function of momentum and centrality.
In Section III the elliptic flow and the ratio of quark to gluon $\Rapt$
are discussed together with their correlation. In Section IV the impact of a realistic
equation of state is presented. In the last Section we summarize the conclusions from the present study.

\section{Modelling the jet quenching}

Our modelling of the jet energy loss is based on the very widely used adiabatic-like approximation 
for which the jet loses
energy in a bulk medium that is independently expanding and cooling. Therefore the 
jet energy loss is considered as a small perturbation of the bulk dynamics. This is
essentially the main assumption in the model till now developed, even if the higher energy at LHC 
should make this assumption inadeguate due to the significant amount of energy that is contained in the high $p_T$ partons intially produced.

The main ingredients of our model can be easily sketched in the following way.

{\it Initial Conditions --} The parton distributions are calculated in the NLO pQCD scheme:
They are parametrized by power law functions as:

\begin{equation}
\frac{dN_f}{d^2p_T}=\frac{A_f}{\left(1+p_T/B_f\right)^{n_f}}
\label{parton-dist}
\end{equation}

with $f=q,\bar q, g$. The transverse momentum $p_T$ is in unit of GeV and the value of the parameters 
$A_f, B_f, n_f$ is given in Table I and are taken from Ref.\cite{Liu:2006sf}. Such a choice is driven mainly
by the intention to make a direct contact and comparison with the several works discussing 
jet quenching with the
flavor conversion \cite{Liu:2006sf,Liu:2007zz,Liu:2008zb,Liu:2008bw,Fries:2009ez}.

\begin{table}
	\centering
		\begin{tabular}{c c c c}
		\hline\hline  
		      & A[GeV]& B[GeV]& $n_f$ \\ [1ex] \hline
			g   & 1440  & 1.5  &   8.0 \\ [0.5ex]
			q   & 670   & 1.6  &  7.9 \\
  $\bar q$  & 190   & 1.9  &  8.9  \\ [1ex]
  \hline\hline
		\end{tabular}
	\caption{Parameters for initial parton minijet distribution given in Eq.(\ref{parton-dist}) at
midrapidity for Au+Au at $\sqrt{s_{NN}}$= 200AGeV. Taken from Ref. \cite{Liu:2006sf}.}
	\label{tab}
\end{table}
 As regards the parton distribution in space coordinates, it scales with the number of binary nucleon
collision $N_{coll}$ according to the standard Glauber model \cite{Glauber:2007a,Nardi:2001}.
The initial conditions for the bulk medium are described by the density profile $\rho({\vec r},z,\tau)$ 
that in the longitudinal direction evolves according to the Bjorken expansion 
at the velocity of light. 
The initial transverse density profile is instead proportional to the standard Glauber model participant
distribution:
\begin{eqnarray}
\rho_{part}(\vb, \vr)=t_A(\vr) \left[1-e^{-\sigma_{NN}\, t_A(\vb-\vr)}\right]\nonumber\\
+t_B(\vb -\vr)\left[ 1-e^{-\sigma_{NN}\, t_A(\vr)} \right] 
\end{eqnarray}
with $\sigma_{NN}= 42$mb, $\vb$ the impact parameter vector and $t_A$ the nuclear thickness function
normalized to the number of nucleons and
given by:
\begin{equation}
t_A(\vr)=\int_{-\infty}^{+\infty}dz \rho_A(\vr, z)
\end{equation}
where $\rho_A(\vr)$ is the nuclear density that we have taken to be a Wood-Saxon (WS) with
radius $R= 6.38 fm$ and thickness $a=0.535 fm$.
The total number of participants is therefore 
\begin{equation}
 N_{part}(b)=\int d^2r \rho_{part}(b, \vr)
\end{equation}

Beyond the Glauber density profile we have also considered a
simplified sharp elliptic (SE) shape with the $\vec x$ and $\vec y$ axis adjusted to reproduce the same
eccentricity of the Glauber model at each impact parameter.
The SE has been used for the description of the bulk in several jet quenching modeling \cite{Liu:2006sf,
Liu:2007zz,Liu:2008zb,Gyulassy:2000gk}.

{\it Bulk Density Evolution --} For the sharp elliptic shape the density space-time evolution is given by:

\begin{eqnarray}
	\rho(x,y,\tau)=\frac{1}{\tau A_T(\tau)}\frac{dN}{dy_z}
\Theta\left(1-\frac{x^2}{R^2_x(\tau)}-\frac{y^2}{R^2_y(\tau)}\right)
\label{dens-sharp}
\end{eqnarray}

where $A_T=\pi R_x^2 R_y^2$ is the transverse area of the evolving fireball and $R_x,R_y$ is the length
of the two axes of the ellipse. These can in general evolve according to an expansion at constant acceleration
or constant velocity:
\begin{eqnarray}
	R_x(\tau)=R_{xo}+v_T\tau+\frac{1}{2}(a_T+\epsilon_a)\tau^2\nonumber\\
	R_y(\tau)=R_{yo}+v_T\tau+\frac{1}{2}(a_T-\epsilon_a)\tau^2
	\label{rxry}
\end{eqnarray}

with $v_T$ the transverse velocity, $a_T \pm \epsilon_a$ the acceleration that through $\epsilon_a$ can
be taken to be different between the x and y direction in order to simulate the
anisotropic azimuthal expansion that reduces the eccentricity with time.
We have used a typical value of $a_T=0.08$ fm$^{-1}$ that generates a final radial flow $\beta=0.4$
with $v_T=0$, $\epsilon_a= 0.04$ fm$^{-1}$ with a slight dependence on the impact parameter of the collision.
However the sensitivity of the jet quenching on these parameters is quite limited.

For the case with the Glauber profile the evolution of the local density is similarly given by  
\begin{eqnarray}
	\rho(x,y,\tau)=\frac{1}{\tau A_T(\tau)}\frac{dN}{dy} P_{eff}(x,y,\tau)
\label{dens-ws}
\end{eqnarray}
where respect to Eq.(\ref{dens-sharp}) the theta function is substituted by the profile function
$P_{eff}(x,y,\tau)$ and $A_T$ is now the effective area given by the space integral of
the profile function:
\begin{eqnarray}
P_{eff}(x,y,\tau)=\frac{N_{part}(x,y,\tau)}{N_{part}(0,0,\tau)}\nonumber\\
A_T =\int\int dx dy P_{eff}(x,y,\tau)
\label{prof-ws}
\end{eqnarray}
where $N_{part}(x,y,\tau_0)$ is given by the Glauber model while the time dependence is determined
by the expansion in the (x,y) plane according to Eq.(\ref{rxry}).

{\it Energy Loss --}
The aim of our work is to explore the consequences of different $T$ dependences of energy loss on the
observables, hence we
employ various schemes for the energy loss. However to make a connection to the large amount of effort to
evaluate gluon radiation in a pQCD frame we will use also the GLV formula at first order in the opacity
expansion \cite{Gyulassy:2000gk,Gyulassy:2003mc}:

\begin{eqnarray}
\frac{\Delta E(\rho,\tau,\mu)}{\Delta \tau}=\frac{9\pi}{4}C_R \alpha_s^3 \rho(x,y,\tau)\,
\tau log\left(\frac{2E}{\mu^2\tau}\right)
\label{eq.glv}
\end{eqnarray}

where $C_R$ is the Casimir factor equal to 4/3 for quarks and 3 for gluons, $\mu=g T$ is the 
screening mass with $g=3$ in agreement with lQCD results \cite{peshier:jpg35} and $\tau$ is the time minus
the initial time $\tau_i=0.2 \, fm$. Furthemore considering massless partons at midrapidity $E=p_T$.
There are corrections to Eq.(\ref{eq.glv}) coming from higher order
that can be approximately accounted for by a rescaling Z factor of the energy loss. However this not really
relevant for the
objectives of the present work because we will renormalize the energy loss in order to have the
observed amount of suppression $\Rapt$ for central collisions, see next Section. 
In fact our purpose is to study how other variables can change once $\Rapt$ has been fixed
to experimetal data for central collisions. 

Usually in the GLV, as well as in other approaches, the temperature evolution of the strong
coupling $\alpha_s$ is discarded. We will consider the impact of such a dependence
to understand the amount of $T$ dependence that can come simply from the asymptotic freedom
in a pQCD approach.
The scale dependence of the strong coupling can be writte as:

\begin{eqnarray}
\alpha_s(Q^2)=\frac{4\pi}{\beta_0\, ln(-Q^2/\Lambda_{QCD})}
\label{eq-alpha}
\end{eqnarray}

where $\beta_0=11-\dfrac{2}{3}N_f $ and the thermal scale $Q^2=(2 \pi\,T)^2$ which allows to get
a correct behavior of the screening mass $\mu$ on the energy scale \cite{peshier:jpg35}. 

\begin{figure}	
\includegraphics[width=6.5cm]{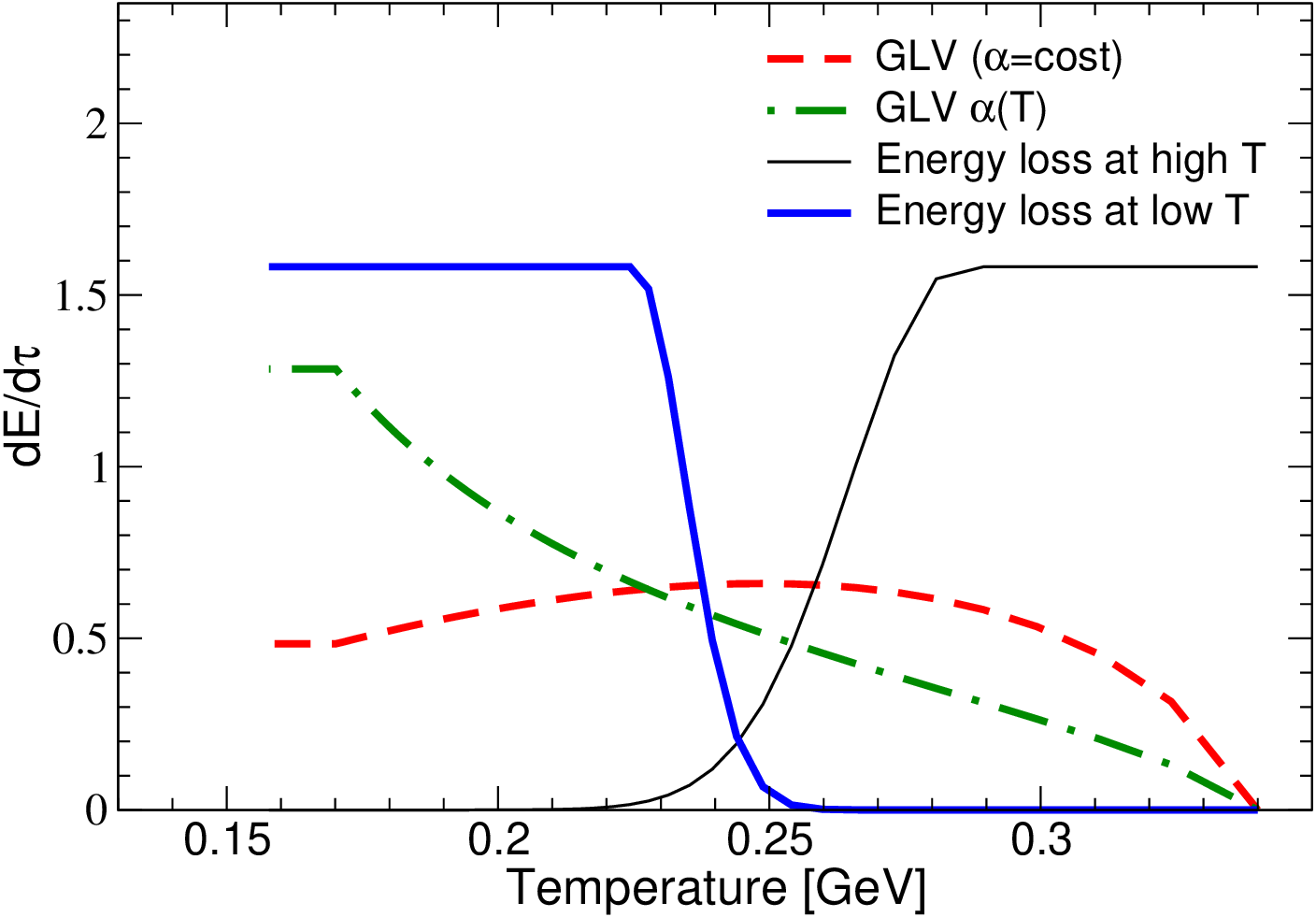}
\caption{Temperature dependence of the energy loss for a parton at $p_T= 10$ GeV. 
By dashed and dot-dashed line the GLV energy loss with constant and T dependent $\alpha_s$ coupling (see
text). The thick line is the case in which the energy loss
take place only closer to the phase transition and the thin line represents an opposite case
in which the energy loss take place only at high T.}
\label{fig-dedtau}
\end{figure}

In Fig. \ref{fig-dedtau} we show by dot-dashed and dashed lines the temperature dependence of the
energy loss
for the GLV with a dependence of the coupling according to Eq.(\ref{eq-alpha}) (GLV-$\alpha_s(T)$) and with
a constant coupling $\alpha_s=0.27$ (GLVc). In itself the effect of the asymptotic
freedom that reduces $\alpha_s$ at increasing energy scale, significantly modifies the temperature dependence
of the energy loss. 
However we will see that such effect is not very large once the coupling itself is readjusted
to produce the correct amount of suppression. 

In Fig.\ref{fig-dedtau} we have also shown two other opposite cases for $\Delta E/\Delta \tau$:
the thick line
that shifts the energy loss to lower temperature (hence low density $\rho$ or entropy density $s$)
as suggested in \cite{Pantuev:2005jt,LS_PRL} and the other (thin line) that gives a dominance of
quenching at high $T$, considered here just for comparison respect to the opposite case.
For both $E_{loss}$ we have a standard dependence on the transverse momentum,
$\Delta E(p_T) \sim p_T^\gamma$ with $\gamma=0.6$.

We have plotted in Fig.\ref{fig-dedtau} the $\Delta E_{loss}$ as a function of the temperature, however under
the hypothesis
of local equilibrium for the bulk this is equivalent to a density or an entropy-density dependence
used by other authors. In the model implemented we generally employ the free gas approximation to
relate density, temperature and time. As well known one has:
\begin{equation}
	\rho=\frac{\zeta(3)}{\pi^2}\left(\frac{3}{4} d_{q,{\bar q}}+d_g \right) T^3\sim 4.2\, T^3
\label{dens}
\end{equation}
where the last expression is obtained with $d_{q,{\bar q}}=24, d_g=16$.
From Eq. (\ref{dens}) with an initial temperature $T_0=340$ MeV at $\tau_0=0.6$ fm 
and a transverse area $A_T \sim 90\, fm^2$
one has a $dN/dy \sim 1000$ for $b=3 fm$ corresponding to $0-10\%$
in agreement with standard estimates.
Eq.(\ref{dens}) allows one to evaluate the local temperature from the local density given by
Eq.(\ref{dens-sharp}) for the SE profile and by Eq.(\ref{dens-ws}) for the Glauber WS profile.
For a 1D Bjorken expansion and the SE profile Eq.(\ref{dens}) gives a direct correlation
between density $\rho$, temperature $T$ and time $\tau$:

\begin{equation}
	\frac{T}{T_0}=\left(\frac{\rho}{\rho_0}\right)^{1/3}=\left(\frac{\tau_0}{\tau}\right)^{1/3}
	\label{temp-time}
\end{equation}
with $T_0, \rho_0, \tau_0$ the values at same initial time.
By mean of Eq.(\ref{temp-time}) we can relate the $\Delta E/\Delta \tau$ represented by the
thick solid line in Fig. \ref{fig-dedtau} to the delayed energy loss employed by Pantuev
\cite{Pantuev:2005jt}. In particular with our parameters from Eq.(\ref{temp-time}) our
thick solid line corresponds to a delay of about 1.8 fm close and even less extreme
than the one in Ref. \cite{Pantuev:2005jt}. One can also notice that 
for some observables already the less extreme GLV$-\alpha_s(T)$ can give results
similar to the low T energy loss, see Fig. \ref{v2_rigi} and \ref{fig-Rapporto-deltaE}.

{\it Hadronization --}  
The final step of the model is the hadronization by independent fragmentation. 
The parton distribution after the jet quenching are employed to evaluate the hadron spectrum
by independent jet fragmentation using the AKK fragmentation functions $D^H_f(x,Q^2)$ which
give the probability that a hadron $H$ is formed from a parton of flavor $f$.
The final hadronic spectrum is obtained from

\begin{eqnarray}
	\frac{dN_H}{d^2p_T}=\int_0^1 dx \, x \,\sum_f \frac{dN_f}{d^2p_T} D^H_f(x,Q^2)
\end{eqnarray}
where $x=p_T^H/p_T^f$ is the fraction of the $f$ parton carried by the hadron H and $Q=p_T^f/2$ is the
energy scale.
At RHIC there have been several evidence that while hadronization by independent fragmentation is able to 
describe proton-proton spectra at $p_T \geq 2$ GeV, in Au+Au collision there are non-perturbative
effects like quark coalescence that modify hadronization at least up to $p_T \sim 5-6$ GeV
\cite{Fries:2008hs,Greco:2007nu}. 
We do not include any hadronization by coalescence with the present work 
therefore all of the following results have to be considered reliable only for $p_T \geq 5$ GeV.

\section{Application of the model to evaluate $\Rapt$}
\begin{figure}	
\includegraphics[width=6.5cm]{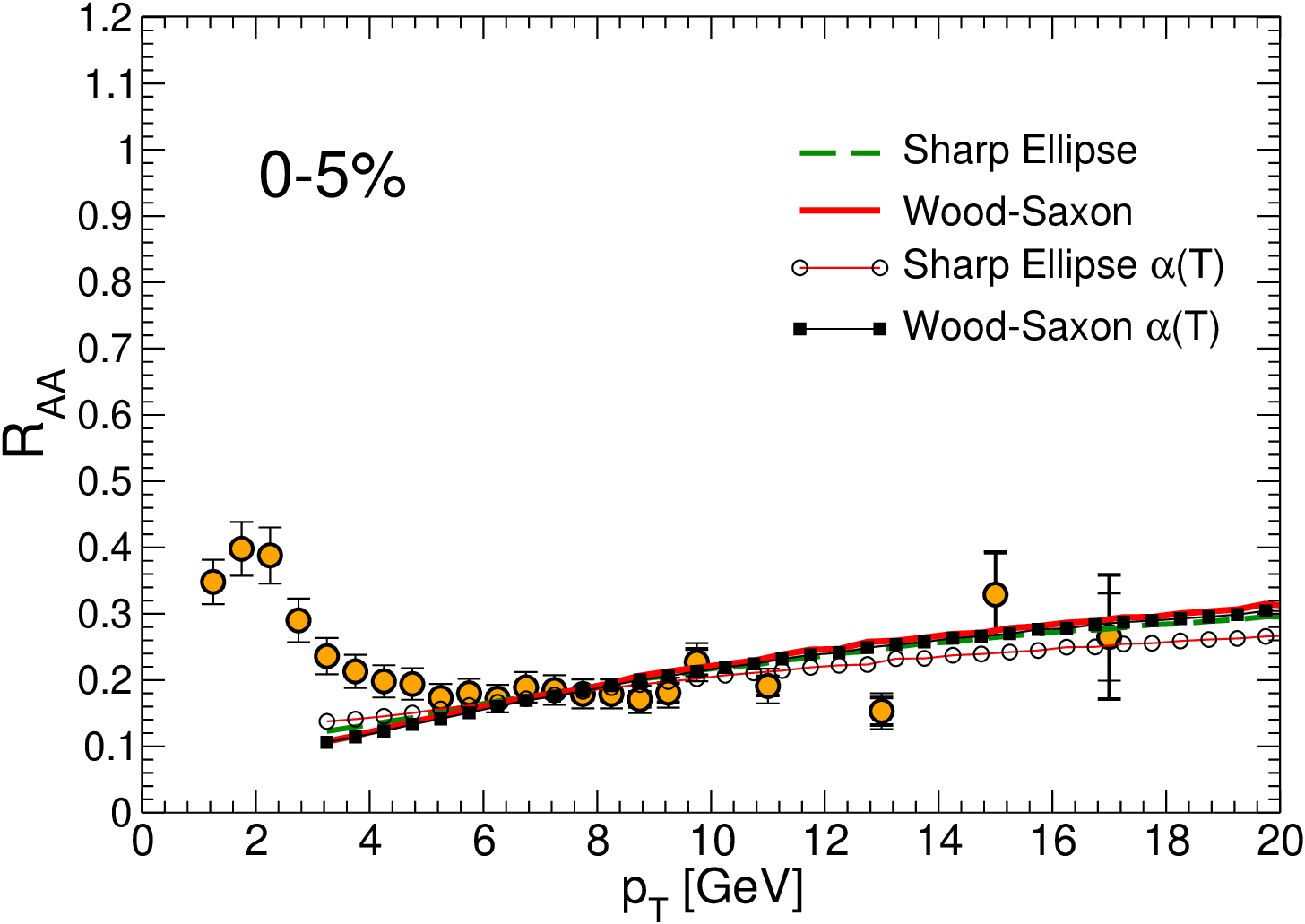}\\
\vspace{1.0cm}
\includegraphics[width=6.5cm]{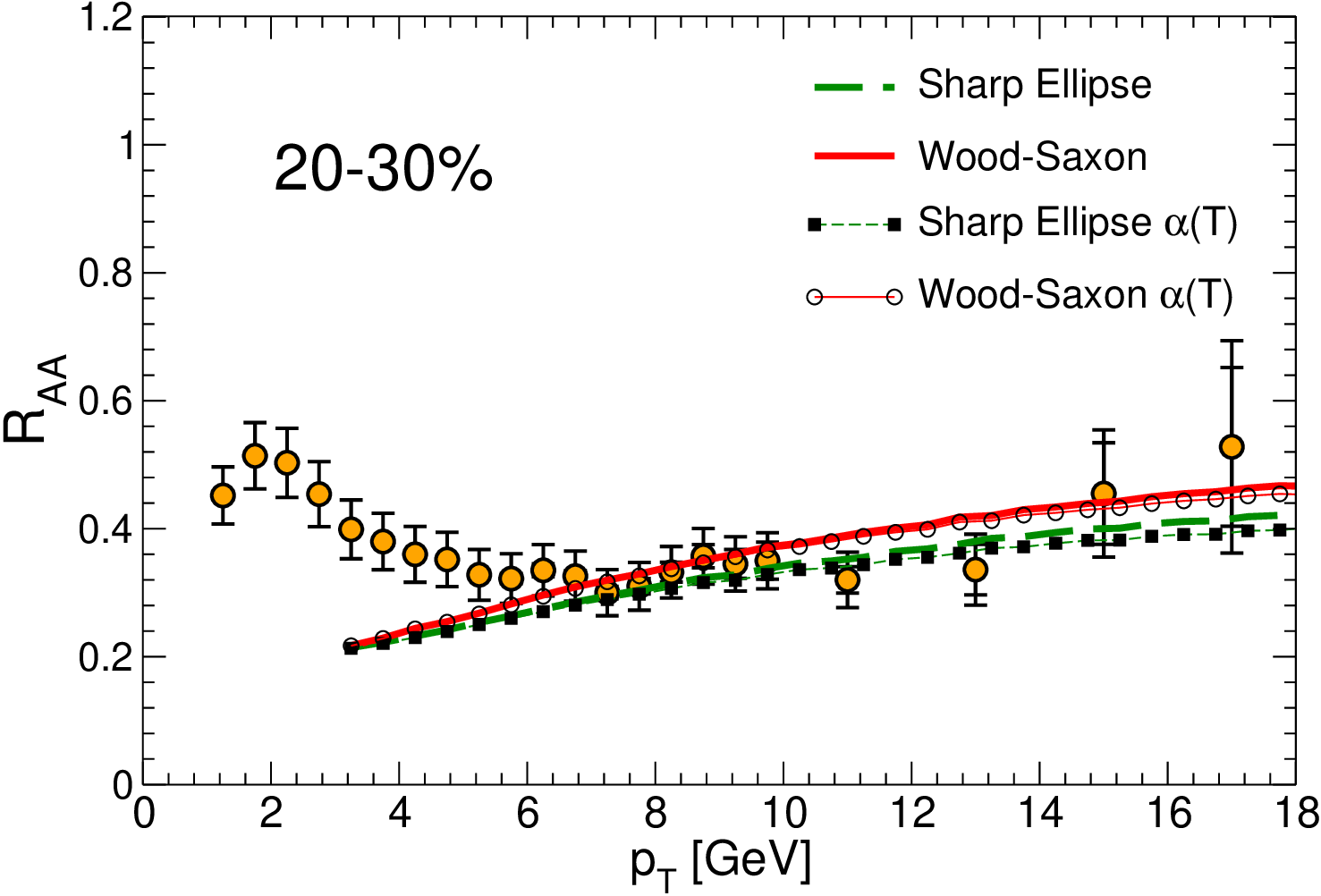}\\
\vspace{1cm}
\includegraphics[width=6.5cm]{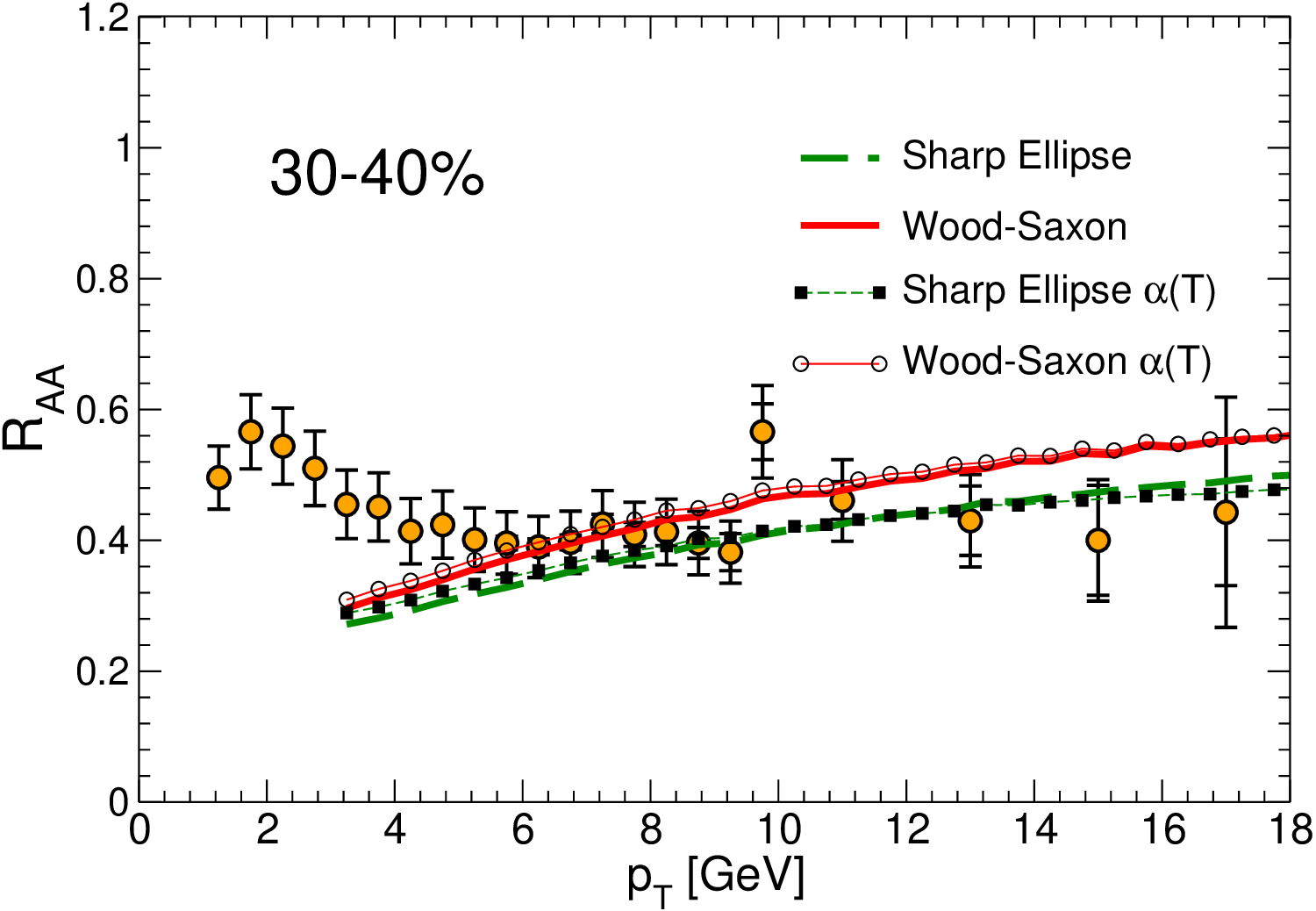}
\caption{Nuclear Modification Factor as a function of the transverse momentum $p_T$ in
$Au+Au$ at 200 AGeV for different
centralities. The circles are the experimental data taken from Ref.\cite{Phenix_more}. The calculation using the GLV
energy loss are shown by dashed and solid line for a Sharp Ellipse profile
and a Wood-Saxon profile respectively. The squares and open circles refer to
GLV$-\alpha_s(T)$ 
for a SE and WS profiles respectively.}
\label{fig-RAApt-centr}
\end{figure}

The amount of quenching  is quantified by comparison of the inclusive spectra $d^2 N^{AA}/dp_td\eta$ in
ion-ion(AA) collision to a nucleon-nucleon(pp) reference $d^2 \sigma^{NN}/dp_td\eta $ via the Nuclear
Modification Factor $R_{AA}(p_t)$. :

\begin{equation}\label{eqn_raa}
R_{AA}(p_t) \equiv \frac{d^2 N^{AA}/dp_td\eta}{T_{AA}\cdot d^2
\sigma^{NN}/dp_td\eta }
\end{equation}

with $T_{AA}$ the nuclear overlap function which scales up single NN cross section to AA according to expected
number of binary NN collisions {\em without} modification. Thus a $R_{AA}$ smaller(larger) than unity means
suppression(enhancement) due to medium effect. At RHIC this ratio at large $p_t> 6 GeV$ has been measured to
be
nearly constant around avalue of $0.2$ for the most central AuAu collisions, see Fig.
\ref{fig-RAApt-centr} (upper). 
In our model the $\Rapt$ can be calculated simply from the ratio of the spectra before and after 
quenching.

We have applied our modelling of the jet quenching for Au+Au collisions at 200 AGeV. 
We use standard initial conditions that for the most central collision bin, $b=3 fm$, 
are given by a $dN/dy=1000$ and a maximum temperature of the bulk $T_0=340$ MeV 
at $\tau_0=0.6 \, fm$, as usually done to describe
the bulk in hydrodynamics and transport approaches \cite{Kolb:2003dz,Greco:nett}.
The results are shown in Fig. \ref{fig-RAApt-centr} and are performed for the
two geometries described above: Wood-Saxon profile (WS) and Sharp Elliptic shape (SE).
The GLV formula, Eq.(\ref{eq.glv}), is used with a coupling $\alpha_s=0.27$ but rescaled by a $Z=0.45$
 factor \cite{multi_gluon_fluct} that accounts for higher
order effect.  Here it has been chosen to reproduce the data at $p_T=6$ GeV for the most
central selection $0-5\%$. From the WS profile to the SE one has to decrease by about
$15\%$ the $Z$ normalization factor due to lack of surface where the quenching is smaller.
However both values are well inside the uncertainty in the $\alpha_s$ strength of the in-medium 
gluon radiation
due to the uncertainty in the perturbative expansion and in the validity of the expansion itself.
However as said in the introduction our purpose is not to constrain the total amount of
quenching due to gluon radiation. Our methodology is to fit the $\Rapt$ in order to 
fix the correct amount of total quenching with the aim of exploring the effect
of different geometries and especially different temperature ($\rho$ or $s$) 
dependences of the energy loss on other observables. 

\begin{figure}	
\includegraphics[width=7cm, height=5cm]{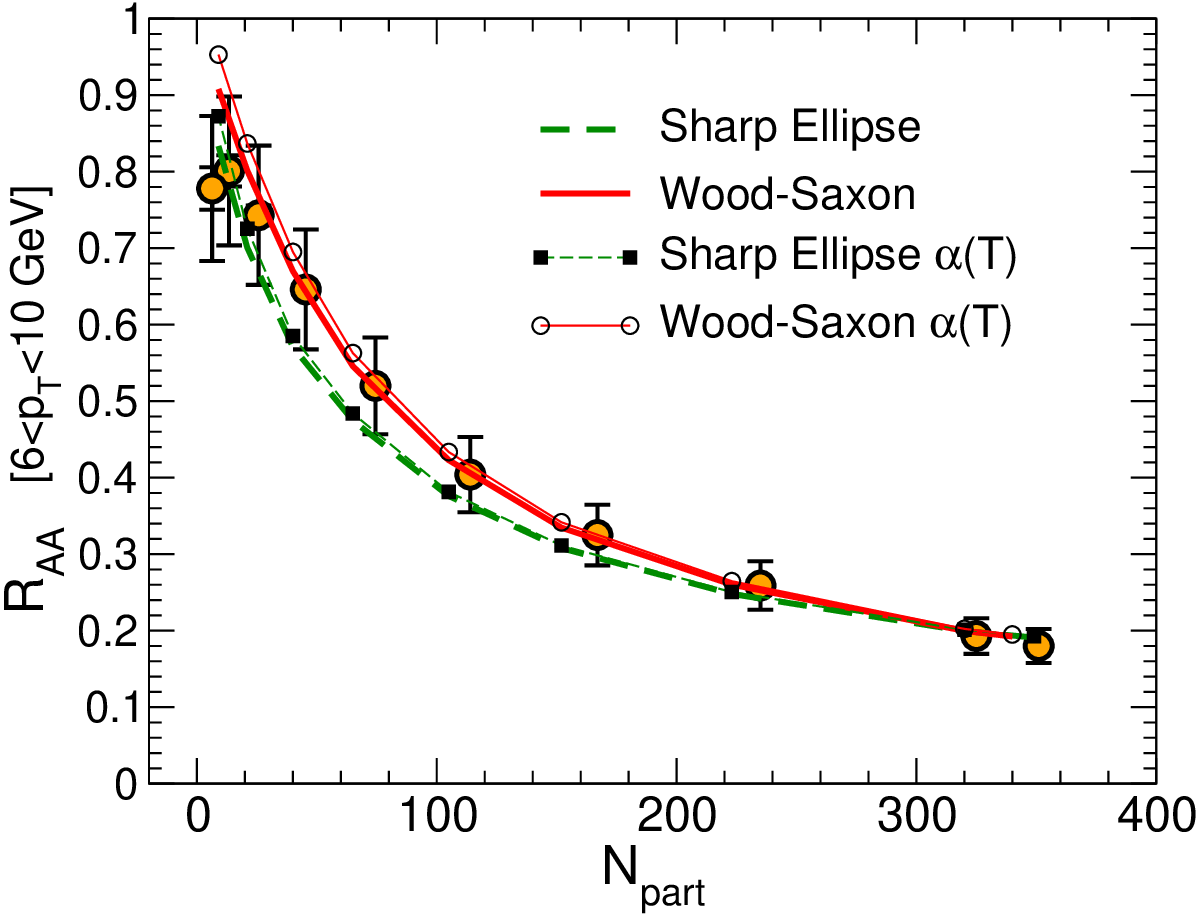}
\caption{Nuclear Modification Factor as a function of the number of participant 
in $Au+Au$ at 200 AGeV for two density profiles of the bulk
matter: Glauber from Wood-Saxon nuclei profile (solid line) and Sharp Elliptic Shape (dashed line).}
\label{fig-RAA-centr}
\end{figure}

In Fig.\ref{fig-RAApt-centr} we can see that once the amount of quenching is fixed for the most 
central collisions both the dependence on $p_T$ and on centrality are correctly predicted with a 
GLV formula for quenching. Of course such a result has been obtained with other models of gluon radiation
like BDMPS, ASW, AMY, DLGV \cite{bdmps,asw,amy,dglv}. Our purpose here was simply to show that our model is
also able to reproduce
the $\Rapt$ which can be considered as the minimum requirement. However our result shows also that 
at the level of $\Rapt$ both a realistic density profile like the WS or the Sharp Ellipse profile
can describe the data reasonably well. 
We have seen that the main reason behind the similar final result between the two different geometries, 
WS and SE, relies on the compensation between two effects. In fact for SE one has a non realistic
uniform density but on the other hand the minijet are also distributed uniformly.
So the uniform density profile from one hand overestimates the amount of quenching close to the surface
but from the other underestimates the one in the core of the fireball. In addition in such a modelling the fact that
minijets are uniformly distributed overestimates the amount of minijets leaving the fireball
nearly unquenched. Even if one could suspect that a balance among these effect
should not apriori hold at all centralities our results show that 
the breaking of such a cancellation effects is small. In fact 
regard the $\Rapt$ a simplified model with the SE cannot be discarded.
The same conclusion can be drawn looking also at the $\Ra$ at $p_T > 6$ GeV as a function
of centrality shown in Fig. \ref{fig-RAA-centr}. However, as expected, we see that surface effects leading to
less suppression for a WS geometry are more important for a smaller $N_{part}$.

Furthermore if we consider a GLV formalism with a running
coupling constant $\alpha_s(T)$, GLV$-\alpha_s(T)$, the $\Rapt$ is well reproduced at same level of quality.
Hence looking at $\Rapt$ one is not able to clearly discriminate neither the geometry nor the
temperature dependence of the quenching even if one looks at the evolution with centrality.
Therefore in agreement with Ref.s \cite{Renk:2007mv,Bass:2008rv}, we find that $\Rapt$ carries only a weak
information on the jet 
quenching process, apart of course the total amount of quenching, which in itself is of fundamental
importance and has lead to a first estimate of the average initial gluon density. 

In the next Section we will investigate both the elliptic flow at high $p_t$ and the flavor dependence
of the quenching. In fact both observables are still hardly accounted for quantitatevely
by the present models.

\section{Angular and Flavor dependence of the Quenching}  

The first analysis of jet suppression has shown that it is very difficult to have an agreement 
between models and experiments for the dependence of the $\Ra$ on the azimuthal
angle $\phi$ respect to the reaction plane in non central collisions \cite{Bass:2008rv}.   
Such a dependence arise from the ``almond'' (elliptic) shape of the 
overlap region of two colliding nuclei. In particular for large $p_t> 6\,
{\rm GeV}$ where hard processes dominate \cite{Phenix_more}, partons
penetrating the fireball in different directions lose different amount of energy according to their varying
paths that on average are larger in the out of plane direction. A measure of this effect is provided by the
second Fourier coefficient of the distribution, namely the elliptic flow:

\begin{equation}
 v_2(p_T)\equiv \frac{\int_0^{2\pi}d\phi\, \cos(2\phi)\,[d^2N/dp_td\phi]}
 {\int_0^{2\pi}d\phi \, [d^2N/dp_td\phi]}
\end{equation}

Unexpectedly, measured $ v_2(p_T)$ happened to be considerably larger than what jet quenching models
predicted. 
It was noted in Ref.\cite{Shuryak:2001me} that for very strong quenching only jets emitted from 
the surface of the ``almond shape'' can be observed and the data for $ v_2(p_T)$ are very close to such a
limiting case. However generally three other assumptions are made, namely:
(i) quenching is proportional to  matter density;
(ii) colliding nuclei were approximated  by homogeneous sharp-edge spheres. 
(iii) Only the net rate of energy loss is considered discarding the emission-absorption processes
that generally can lead to an enhancement of the collective flows \cite{amy}.
Studies by Drees,{\it et al} \cite{Drees:2003zh} relaxed the second assumption, with realistic
nuclear shapes, which only made contradiction with data even stronger. A result that is 
confirmed also by the present work.

\begin{figure}	
\includegraphics[width=7.5cm,height=5cm]{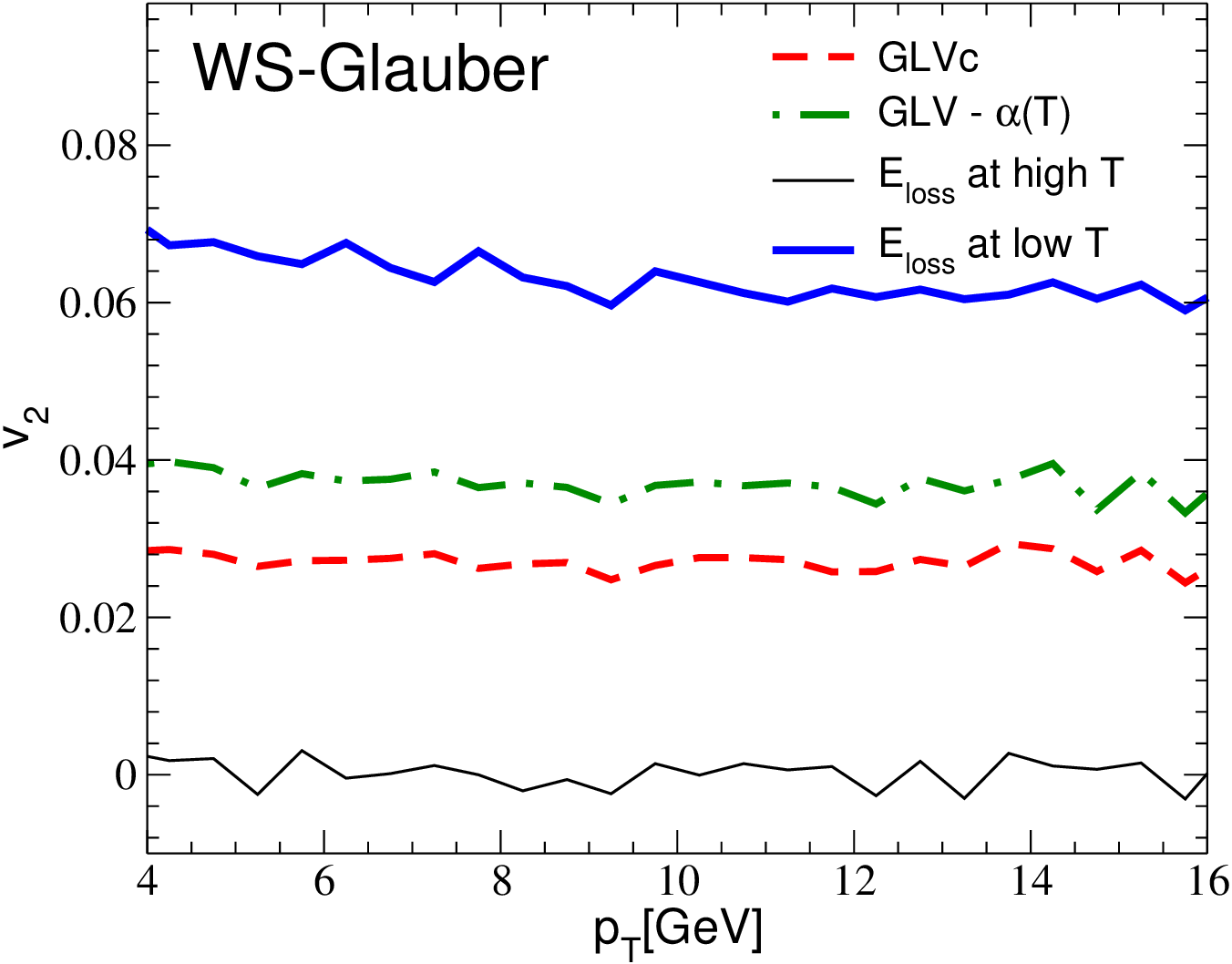}
\caption{Elliptic Flow for the pions coming from quark and gluon fragmentation for the different T dependence
of the energy loss as shown in Fig.\ref{fig-dedtau}. The density profile of the bulk is given by
the Glauber model (see section II).}
\label{v2_wood}
\vspace{0.5cm}
\end{figure}

\begin{figure}	
\includegraphics[width=7.5cm,height=5cm]{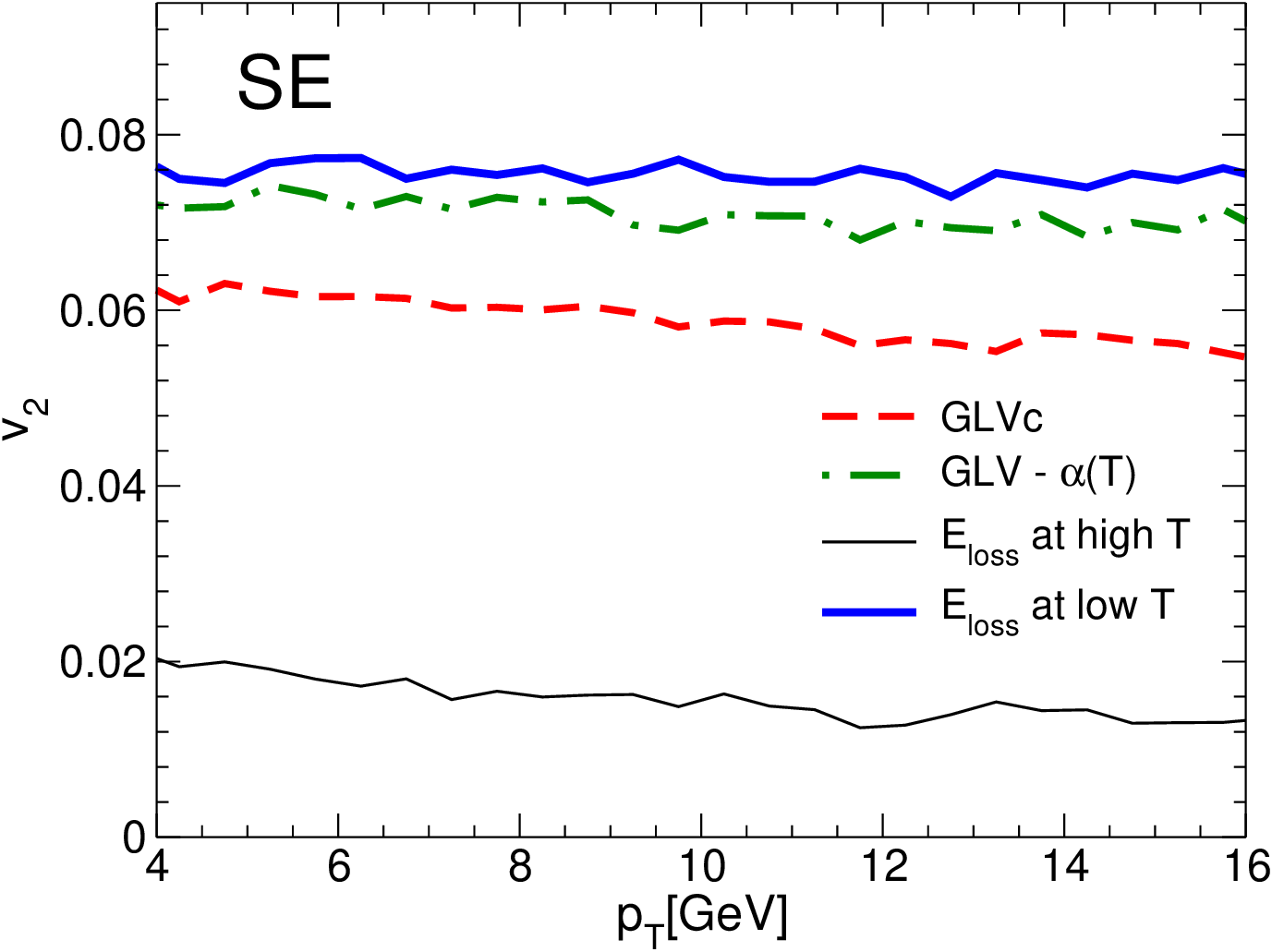}
\caption{Elliptic Flow for the pions coming from quark and gluon fragmentation for the different T dependence
of the energy loss as shown in Fig.\ref{fig-dedtau}. The bulk density profile is a sharp uniform elliptic
shape.}
\label{v2_rigi}
\end{figure}

We will mainly explore the effect of the first assumption by exploring the four different
kinds of energy loss shown in Fig. \ref{fig-dedtau}.
In Fig. \ref{v2_wood}, we show the resulting $\v2$ for pions in $Au+Au$ at $\sqrt{s_{NN}}=200$ GeV and
an impact pararameter $b=7.5$ fm corresponding to a minimum bias condition for which the experimental value 
of elliptic flow is around 0.11 as shown in Fig.\ref{v2_RAA_WS}. 
We can see that even if the amount of total quenching has been fixed to the experimental
value of $\Rapt$ the amount of elliptic flow is still strongly dependent on the 
temperature dependence of the $\Delta E_{loss}$. Specifically from thin to thick
solid line, we see that $v_2$ increases if the $E_{loss}$ is stronger as 
$T\rightarrow T_c$ (thick solid line of Fig.\ref{fig-dedtau}) which means later in the evolution of the QGP,
except for the surface.
In Fig.\ref{v2_rigi} the same quantity is plotted for the case of a bulk density
given by the Sharp Ellipse. Again we find that in agreement with other calculations 
\cite{LS_PRL, Pantuev:2005jt}, the elliptic flow 
is significantly larger if most of the energy loss take place closer to
$T_c$ (thick black line).  If instead the energy loss take place mainly
at high temperature/density (thin black line) the $\v2$ is essentially vanishing. 
The dashed line is the result with the GLVc that represent a case in which
the energy loss is essentially proportional to the density. The dot dashed line is the
GLV-$\alpha_s(T)$ that reweigths the temperature dependence through
the coupling $\alpha_s(T)$ giving an estimate of the effect of asymptotic freedom.
Such an effect is not negligible and for the SE geometry it seems to give already a $\v2$
very close to the much more extreme case rappresented by the thick solid line,
i.e. quenching dominated by the close to $T_c$ region.

The correlation between the temperature dependence
of $\Delta E$ and the amount of elliptic flow developed is clear.
In fact at variance
with $\Rapt$ the $\v2$ has a longer formation time because the minijets have to
explore the shape of the fireball. For example if one consider the extreme case of 
a very strong quenching that takes place on distances much lower than the fireball
size then there would be no elliptic flow (like in the $E_{loss}$ at high $T$ case). 

It seems to be quite likely that experiments are telling us that
quenching is likely to be $not$ proportional to the matter entropy density or temperature, but a
decreasing function of it. This is what is essentially discussed  
in Ref.s\cite{Pantuev:2005jt,LS_PRL} 
%it was realized that if the quenching is dominant in the later
%stage of the QGP phase, i.e. closer to $T_c$, the amount of quenching of $\v2$ increases getting closer to the
%experimental data. However in such a modelling 
where however it was implicitely assumed that the amount of quenching
of quarks and gluons are equal among them and to the hadronic one. Here we relax such assumptions
showing that the temperature (or entropy density) dependence of $\Delta E_{loss}$ modifies not only the
$\v2$ but also the relative amount of quenching of quarks and gluons. This does not want to be
just a more detailed
calculation but is indeed related to another puzzle of the jet quenching phenomena 
observed more recently from the experimental study of the chemistry of the minijet suppression
\cite{Sickles_QM09,raa-flavor}. We discuss it in the next subsection.

\subsection{Quark to Gluon Modification Factor}

The QCD due to its SU(3) Lie algebra gives a factor $C_R=9/4$ larger for the energy loss of
gluons respect to that of quarks. For this reason sometimes it is implicitely assumed that the ratio between
the suppression of the gluons $\Ra(g)$ and the suppression of quarks $\Ra(q)$ is such that
$\Ra(q)/\Ra(g)=9/4$.
From this expectation one would think that the (anti-)protons are more suppressed respect
to pions because they
come more from gluon fragmentation than from quarks fragmentation respect to pions. 
The data at RHIC however has shown that also for such an observable there is no agreement with the
data. In fact even outside the region where coalescence should be
dominant \cite{Fries:2008hs,Greco:2007nu} the protons and the antiprotons appear to be less suppressed than
the pions
and $\rho^0$ \cite{Sickles_QM09,raa-flavor}. 
Again we can see that going beyond the simple amount of quenching given by $\Rapt$ neither the
azimuthal dependence nor the flavor dependence of the quenching appear
to be in agreement with the data. 
We call this open issues the "azimuthal" and the "flavor" puzzle respectively.
We will show that even if $\Ra$ for central collisions is fixed to be $\sim 0.2$ also the
$\Ra(q)/\Ra(g)$ is affected by the temperature dependence of $\Delta E_{loss}$

\begin{figure}	
\includegraphics[width=6.5cm,height=5cm]{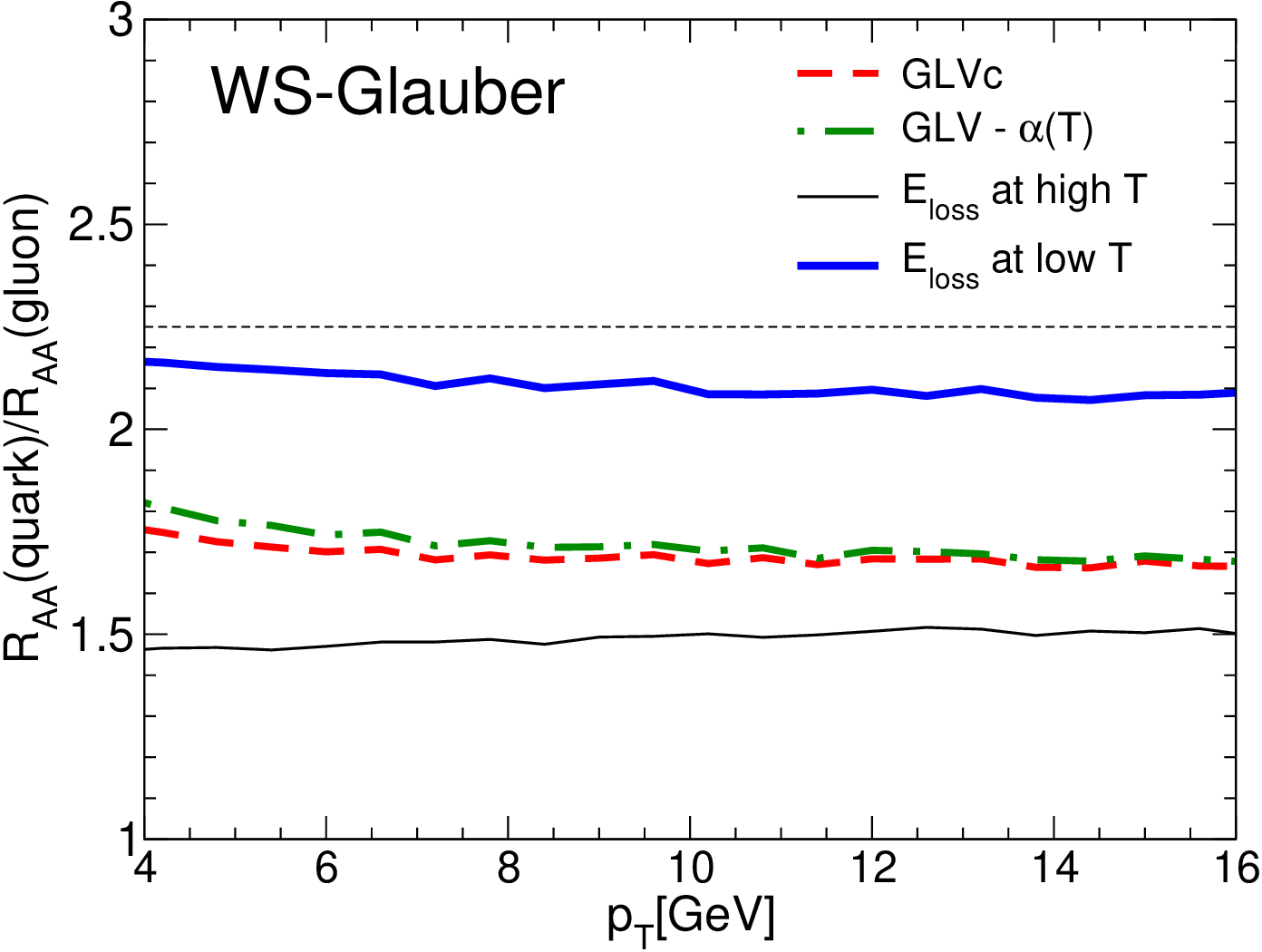}
\caption{Ratio of quark to gluon $R_{AA}$ for the different T dependence of the energy loss as shown in
Fig.\ref{fig-dedtau}. The short-dashed line corresponds to the $9/4$ value. The density profile of the bulk is
that given by the Glauber model.}
\label{rapporto_RAA_wood}
\end{figure}

In Fig.\ref{rapporto_RAA_wood} we show the ratio of the $\Ra(q)/\Ra(g)$ for the WS geometry and the four
different temperature dependences of the energy loss $\Delta E_{loss}$ as in Fig.\ref{fig-dedtau}.
We can see that the standard GLVc energy loss does not give the expected ratio $9/4$ for $\Ra(q)/\Ra(g)$
but a lower value around 1.8 which represents already a non negligible deviation from 2.25.
We can however see that if the energy loss would be strongly T dependent and dominated by
the $T \gtrsim T_c$ region $\Ra(q)/ \Ra(g)$ can increase up to about 2.3 while
oppositely if it is dominated by the high temperature region (thin solid line)
the $\Ra(q)/\Ra(g)$ can become as small as 1.5. On the other hand our study of the elliptic flow,
as well as previous studies,
show that an energy loss that increase with $T$ would generate a tiny $\v2$ very far from the observations.
Such an effect is totally discarded in 
ref.s\cite{Pantuev:2005jt,LS_PRL} that neglect the different quenching of quarks and gluons.
In other words while the indication of a $\Delta E(T)$ increasing as $T\rightarrow T_c$
is confirmed also by our model in order to reproduce experimental data for $v_2$, we have spot that
this would lead to 
a larger $\Ra(q)/\Ra(g)$ ratio increasing the disagreement with the $\Rapt$ observed
for the various hadrons like $p,\bar p, \pi$ that show the $\Ra$ of baryons less suppressed
respect to the pionic one.

\begin{figure}	
\vspace{0.4cm}
\includegraphics[width=6.5cm,height=5cm]{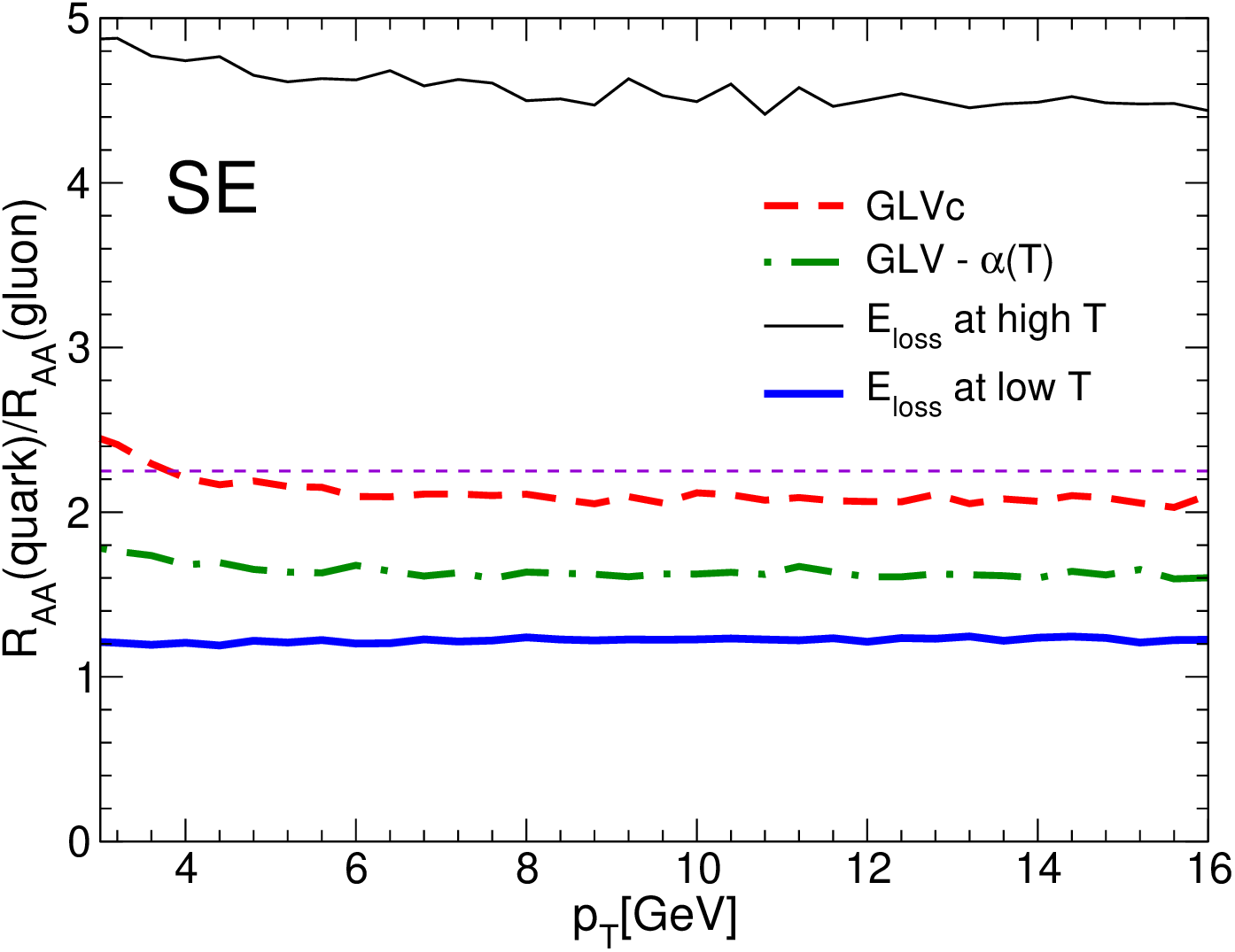}
\caption{Ratio of quark to gluon $R_{AA}$ for the different T dependence of the energy loss as shown in
Fig.\ref{fig-dedtau}. The short-dashed line corresponds to the $9/4$ value. The density profile of the bulk is
that of a sharp elliptic shape.}
\label{rapporto_RAA_SE}
\end{figure}

In this respect we spot also another potential problem that can arise for 
simplified fireball modelling when extreme $\Delta E(T)$ are considered and more
exclusive observables are investigated.
More explicitly we refer to the Sharp Ellipse (SE) fireball that neglects the surface
of the fireball. 
In Fig. \ref{rapporto_RAA_SE} we show that the results have an opposite pattern for the SE geometry 
respect to the realistic WS one. 
The standard GLV energy loss gives approximately the expected ratio of $9/4$ for $\Ra(q)/\Ra(g)$; 
This probably led to associate the $9/4$ factor on $E_{loss}$ to that on $\Ra(q)/\Ra(g)$.
However this ratio is more sensitive to the temperature dependence of the energy loss, in a way that is
exactly 
opposed to that we have seen for the WS geometry. For this kind of geometry (SE) if the energy loss is
stronger at high temperature, the $\Ra(q)/\Ra(g)$ can result to be almost $5$ (thin solid line) on the
other hand, if the enery loss is stronger closer to $T_c$ the $\Ra$ would be just sligthly above 1 (thick
solid line).
This means that the ratio tends to decrease from 5 to 1 if the energy loss is
dominated by the high temperatures ($T \sim 2T_c$) or by the low ones ($T \sim T_c$).

It is instructive to explain the origin of such a strong effect of the density profile,
but before we note that such an un-expected strong effect shows up only when extreme
$\Delta E(T)$ are considered. In fact the standard GLVc $E_{loss}$ is modifed by 
about a $15\%$ moving from a WS-Glauber to a simple SE profile.

\begin{figure}	
\includegraphics[width=7cm, height=5cm]{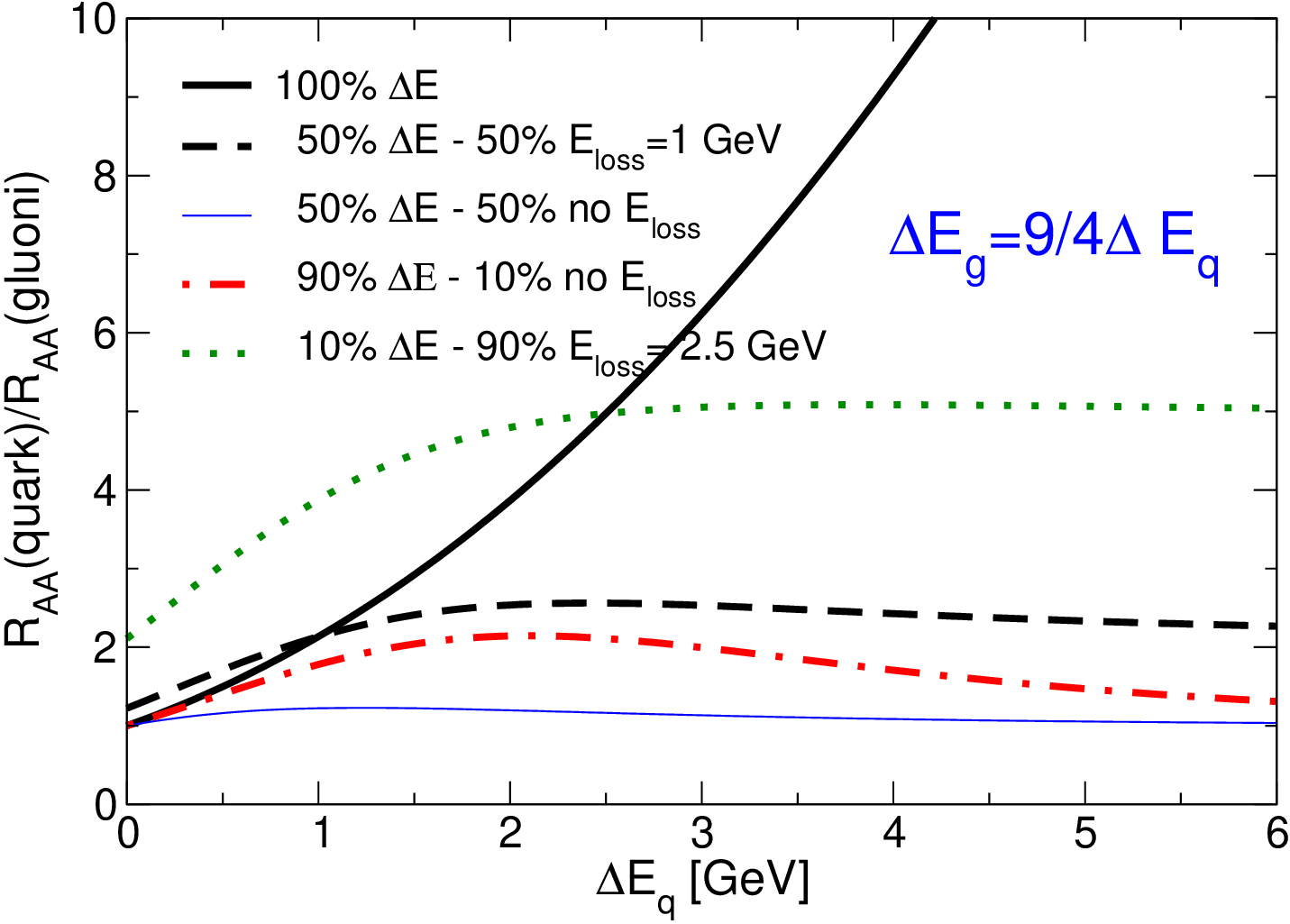}
\caption{Ratio of quark to gluon $\Ra$ as a function of the quark energy loss
$\Delta E_q=4/9 \Delta E_g$ for different models of the jet quenching, see text for details.}
\label{fig-Rapporto-deltaE}
\end{figure}

In order to understand the mechanism behind the determination of the ratio $\Ra(q)/\Ra(g)$ we discuss an
oversimplified case in which all quarks lose the same amount of energy and all gluons lose their energy
according
to $\Delta E_g=9/4 \Delta E_q$.
For such a simple case the spectra after quenching are shifted by a quantity equal to the lost energy.
Quarks that finally emerge with an energy $E_f=p_T$ are those which before quenching had an 
energy $E_i=p_T+\Delta E$. 
So $\Ra$ for quarks is equal to the ratio between the parton distributions in momenta space without quenching
$f(p_T)$ and the quenched one given by $f(p_T+ \eta\,\Delta E)$ 
where $f(p_T)=dN/d^2p_T\,dy$ is given by Eq. (\ref{parton-dist}), $\eta=1$ for quarks
and $\eta=9/4$ for gluons. Therefore the $\Ra$ is

\begin{equation}
 \Rapt =\frac{f(p_T+\eta\,\Delta E)}{f(p_T)}
\end{equation}
and the ratio between quark and gluon nuclear modification factors is 

\begin{equation}
\frac {R_{AA}(q)}{R_{AA}(g)}={\frac{f_q(p_T+\Delta E)}{f_q(p_T)}}\frac{f_g(p_T)}{f_g(p_T+(9/4)\,\Delta E)}
\end{equation}

of course there is no reason why this ratio must be $9/4$
and we can also see that even without the $9/4$ factor there can be a $\Ra(q)/\Ra(g)$ that is not one. 
In Fig.\ref{fig-Rapporto-deltaE}  is shown the dependence of the ratio on $\Delta E$ for partons 
with $p_T=10\, GeV$  and we can see that $\Ra(q)/\Ra(g)$ quickly increase with $\Delta E$ even if the ratio between the quark to gluon energy loss is
$9/4$. The ratio is about 4 for a $\Delta E \sim 2$ GeV/fm. We will see that
this observation is  fundamental  to understand the relation between $\Delta E_{loss}$ and the ratio $\Ra(q)/\Ra(g)$. 
In this very simplified model the $\Ra(q)/\Ra(g)$ would be very large for
an average energy loss $<\Delta E_q>= 4 GeV$ typical of central collisions
where $\Rapt \sim 0.2$. From Fig.\ref{fig-Rapporto-deltaE}
we can see that this would give a $\Ra(q)/\Ra(g) \sim 8$.

One can move toward a minimal realistic model
distinguishing among partons two classes of particles:
those that undergo a large quenching and those that lose no energy or better a small
amount of energy, usually associated with the minijets generated at the surface.
Let's consider a case in which $50 \%$ partons lose an energy 
$ \Delta E_q= 1 GeV= 4/9 \,\Delta E_g$ while the other $50 \%$ lose the energy $\Delta E_q$ 
indicated in Fig.\ref{fig-Rapporto-deltaE}. In such a case the evolution of $\Ra(q)/\Ra(g)$
is very different respect to the first case as shown by the much milder increase
of the dashed line that reaches a maximum of about 2.6
and even decrease for a $\Delta E_q> 2$ GeV.

This clearly indicates that because of the rapidly falling distribution
with $p_T$, the determination of the ratio is dominated by those
partons that suffer less energy loss .
If one for example suppose that $50 \%$ of the particles do not lose energy
then whatever is the energy loss of the other partons the $\Ra(q)/\Ra(g)$ will be equal
to one. This is shown by the thin solid line in Fig. \ref{fig-Rapporto-deltaE}.
Indeed already if only a $10\% $ of partons do not lose energy
the ratio stays below $9/4$ whatever is the $E_{loss}$ of the other
$90\%$ of particles and even decrease if those particle lose a large amount
of energy bringing the value again closer to one. 
This is shown by the dot dashed line in Fig.\ref{fig-Rapporto-deltaE} where one
can realize the 
huge impact of the particles that do not lose energy comparing thick
solid and dot-dashed lines which differ only by the fact that a $10\%$
particles do not lose energy. Finally with the dot line in Fig. \ref{fig-Rapporto-deltaE}, 
we show a system in which most of the particles
undergo a quenching with $\Delta E_{q}= 2.5\, GeV=4/9 \Delta E_{g}$. 
This is closer to a case in which most of the particles undergo a similar strong quenching.  
In summary we can say that the $\Ra(q)/\Ra(g)$ is not really directly
linked to the relative amount of quark and gluon energy loss because 
it is largely affected by the way the energy loss is distributed among partons.
In particular once there are minijets that do not suffer energy loss the ratio
$\Ra(q)/\Ra(g)$ gets closer to one, because it is more affected by these minijets.
Hence also a careful treatment of the corona effect would much likely give a 
significant contribution to the determination of the $\Ra(q)/\Ra(g)$.

On the base of the above discussion it is possible to understand
the dependencce of $\Ra(q)/\Ra(g)$ on the temperature dependence $\Delta E(T)$,
seen in Fig.\ref{rapporto_RAA_wood} and \ref{rapporto_RAA_SE} and its
opposite behavior between the WS and SE geometries for the density profile.
In the case of energy loss at low temperature with a SE profile the quenching happens 
at the end of the lifetime 
of the fireball because there is a direct relation between time and temperature, 
see Eq.(\ref{temp-time}). Therefore in the case of the energy loss increasing as $T\rightarrow T_c$
many particles escape without losing energy for SE profile.  
Only the particles in the inner part of the fireball are quenched.
This means that we are closer to the schematic case described by the thin solid line 
for large $\Delta E$ in Fig. \ref{fig-Rapporto-deltaE},
most particles do not lose energy the rest lose a large amount of energy,
and in fact $\Ra(q)/\Ra(g)$ is close to one, see Fig.\ref{rapporto_RAA_SE}.

Instead if the quenching is larger at high temperature all  particles lose energy early, and except for those
very close to the surface most of particles $\Delta E \gtrsim 2.5$ GeV. 
In this case there are essentially no particles that do not lose
energy. So we are closer to the case described by the dotted line for which
$\Ra(q)/\Ra(g)$ is about 5 to be compared to thin solid line of Fig.\ref{rapporto_RAA_SE}.

In the case of WS geometry in general there is no direct relation between
time and temperature because on the surface one will have already at early times
a low temperature.
Therefore in this case an energy loss dominated by high temperatures means that 
the quenching is large only on the inner part of the fireball while particles
in the surface lose a small amount of energy. We are close to the case described
by the dot dashed line in Fig.\ref{fig-Rapporto-deltaE} and the $\Ra(q)/\Ra(g)$
is in fact about 1.5. It is difficult to reach one because due to the
density profile anyway all the particles lose at least some finite energy.
Instead if the quenching takes place mainly at low temperature ($T \sim T_c$), 
this means that energy loss is strong in a layer on the surface of the fireball, 
and because all particles must go through this layer at
some time, all of them lose a large amount of energy.
We understand that this is very different, essentially the opposite, respect to SE case.
In the latter case a $\Delta E(T)$ with a maximum at $T \sim T_c$ means that most of the
particles escape due to the direct time-temperature relation that is not dependent on space. 
For WS-Glauber profile such a case means just the opposite, i.e. all the particles lose a similar amount of
energy. This is essentially
the reason underlying the opposite trend seen in Fig. \ref{rapporto_RAA_wood} and
\ref{rapporto_RAA_SE}.
We finally notice that the importance of such details emerge only if extreme energy losses
T-dependence are considered. 

The conclusion of this study is twofold:
i) If one tries to reproduce the large value of elliptic flow using a type of energy
loss that increase with decreasing temperature there is a simultaneously increase
of the $\Ra(q)/\Ra(g)$ enhancing the discrepancy respect to the observed
``flavor'' dependence of the suppression that seems to prefer a ratio close or even smaller
than one \cite{Sickles_QM09}.

ii) When peculiar energy dependences are considered the specific density-temperature
profile can become very important for a quantitative evaluation of the observables.

These results bring us to make two important observations: 
1) The SE profile is able to describe the observed $\Rapt$, but it is inadeguate to reproduce the ratio 
$\Ra(q)/\Ra(g)$, 
and furthermore, it cannot be used even for a qualitative analisis of this ratio because it gives opposite
results to the more realistic
WS profile.
2) In Fig.s \ref{v2_wood} and \ref{v2_rigi} we have shown that an enhancement of the energy loss near $T_c$
increases the elliptic flow 
toward an agreement with the experimental data. On the other hand such a $\Delta E(T)$
is associated to an enhancement of $\Ra(q)/\Ra(g)$ in apparent disagreement with the data
\cite{raa-flavor}.

\section{Correlation between $R_{AA}(q)/R_{AA}(g)$ and elliptic flow}

To solve what we have called the "flavor puzzle" inelastic collisions 
that cause a change of the flavor has been invoked \cite{Liu:2006sf,Liu:2007zz,Liu:2008zb,Liu:2008bw}. 
Such a process would at the end produce a net conversion of quarks into gluons. Hence a
decrease of gluon suppression respect to the direct suppression and an increase of the quark one. 
\cite{Liu:2006sf}. 
In Ref.\cite{Liu:2006sf} it has been calculated the conversion rate of a 
quark jet to a gluon jet and vice versa due to two-body scatterings. An enhancement factor $K_c=4-6$
that accounts for non-perturbative effect is needed to produce a nearly equal suppression of
quarks and gluons. This is not an unreasonable enhancement factor considering
that at our energy a $K\sim 4$ is necessary also to have the right minijet initial distributions in pp if one starts from a simple second order pQCD calculations.
We have included the flavor conversion mechanism in our code in a fashion similar
to Ref. \cite{Liu:2006sf}. 
In order to check our code and to have a direct link to the previous calculations of Fries, Ko and Liu
we have used the same density profile (i.e. a SE profile), same conversion rate and $\Delta E_{loss}$
derived by them at leading order in the pQCD expansion.

\begin{figure}
\includegraphics[width=7.cm,height=5.cm]{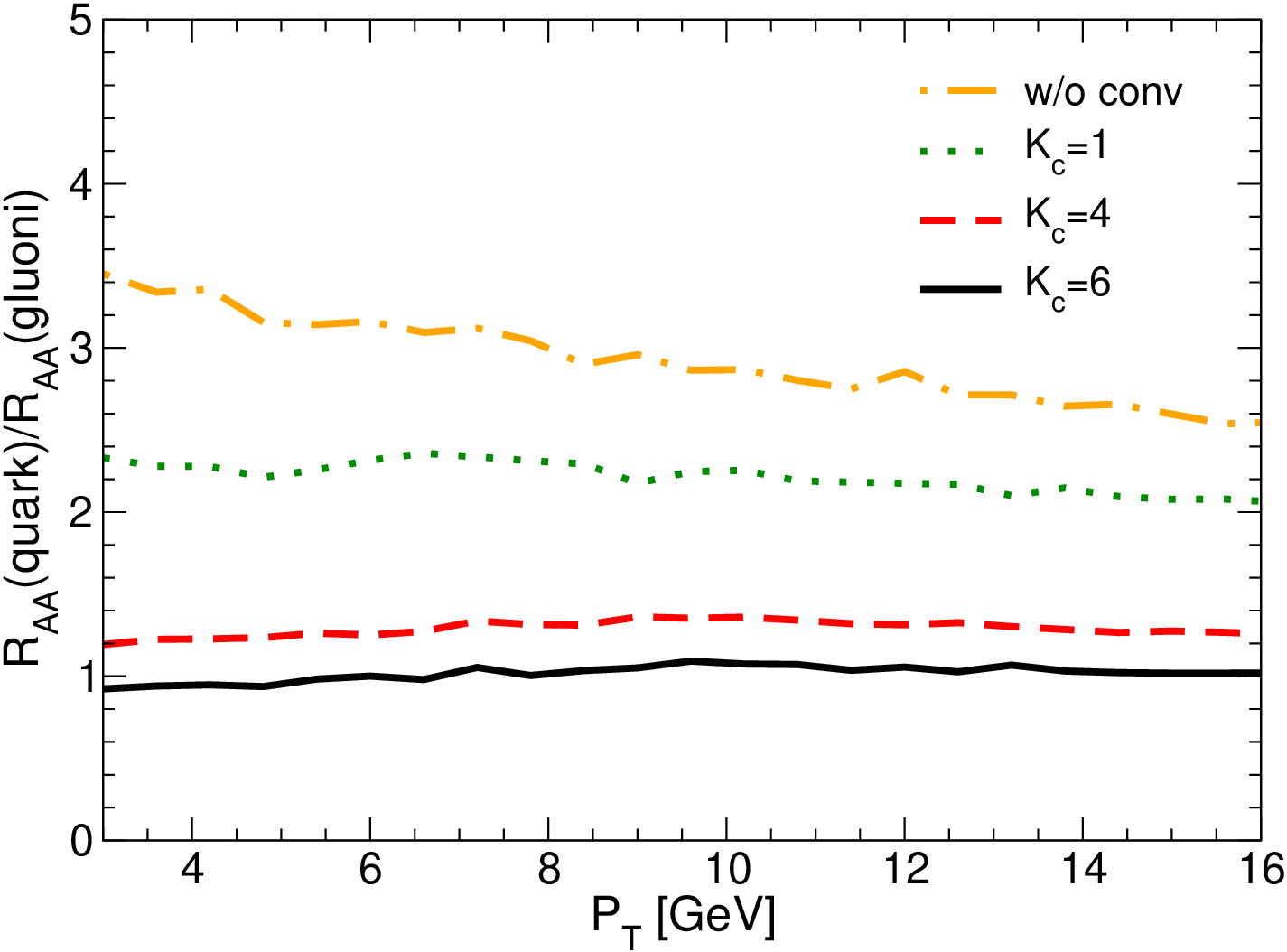}
\caption{Ratio of quark to gluon in $Au+Au$ collisions (b=7 fm) for the energy loss derived in
Ref.\cite{Liu:2006sf} with different factor upscaling the pQCD conversion rate.}
\label{rapporto-Ko}
\end{figure}

\begin{figure}	
\vspace{0.6cm}
\includegraphics[width=7.5cm,height=5.5cm]{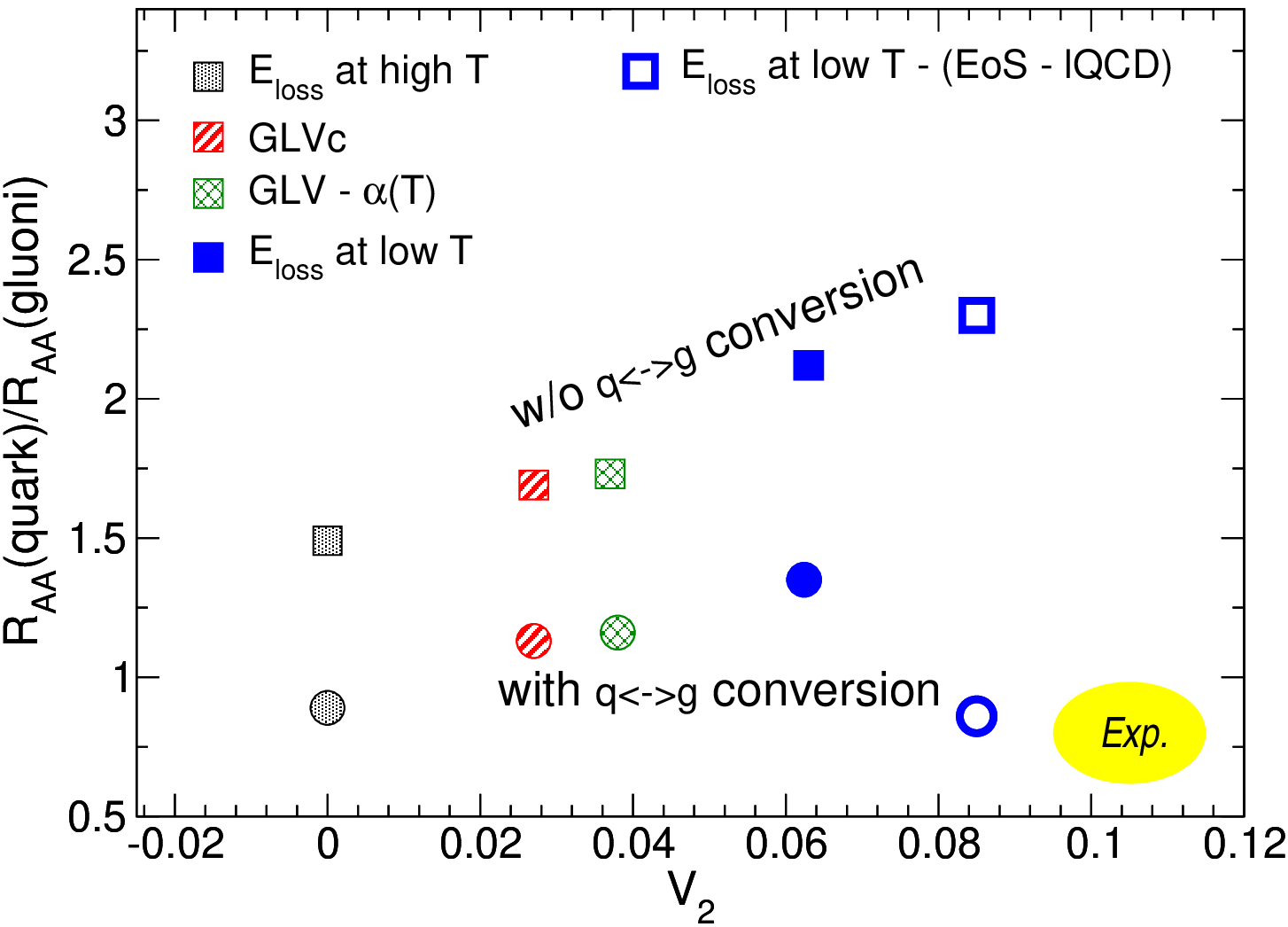}
\caption{Correlation between the $R_{AA}(q)/R_{AA}(g)$ and the elliptic flow of pions ($6<p_T<10$ GeV)
for a WS density profile.
The squares refer to calculation without jet conversion while the triangle are results including jet
conversion with a $K_c=6$ factor. }
\label{v2_RAA_WS}

\end{figure}

In Fig.\ref{rapporto-Ko} we show $\Ra(q)/\Ra(g)$ for different values of $K_c$. Notice that
with the $\Delta E_{loss}$ of Ref.\cite{Liu:2006sf} the ratio $\Ra(q)/\Ra(g)$ 
without conversion is again
different and moreover quite larger than 9/4. However a $K_c \sim 4-6$ is able to reduce that ratio
by about a factor of three making it close to unity.
After this test we have fixed $K_c=6$ and performed the calculation with the four different
energy loss as shown in Fig.\ref{fig-dedtau}. 
In Fig.\ref{v2_RAA_WS} we show directly the $(\Ra(q)/\Ra(g), v_2)$ plot.
This plot manifest a clear but non-trivial correlation between $\Ra(q)/\Ra(g)$ and the $\v2$.
The upper symbols are the results without the jet flavor conversion 
and we can see that such a correlation drives the data far from
the experimental observed values of a $v_2 \sim 0.1$
and an $\Ra(q)/\Ra(g) \leq 1$ (to account for the $\Ra(p+\bar p)>\Ra(\pi^++\pi^-)$ with AKK fragmentation
function).
The lower symbols corresponds to the results including the rate of inelastic collisions and
we can see that this process allows to get closer to the experimental region
because they change the $R_{AA}(q)/R_{AA}(g)$ without modifying the elliptic flow of pions. 
Fig. \ref{v2_RAA_WS} essentially demonstrates that if one tries to reproduce
simultaneously the $\Rapt$ and $\v2$ and $\Ra(q)/\Ra(g)$ a $\Delta E(T)$  increasing
as $T \rightarrow T_c$ is needed but also that in such a case a flavor conversion process becomes 
even more necessary.

However at this point a quantitative study should be performed employing 
a more accurate dynamics for the bulk evolution given by
hydrodynamics or partonic transport theory that
include three-body radiative processes \cite{Fochler:2010wn,Xu:2004mz}. In fact, 
especially if more peculiar energy loss dependence has to be investigated, it is important to have a
quite realistic density and energy density profile as we have discussed above.
Furthermore fluctuations in the energy loss \cite{Gyulassy:2003mc,Wicks:2005gt},
the gain-loss processes \cite{amy} and the elatic energy loss \cite{Qin:2007rn,Auvinen:2009qm}
should be included because
they can give correction to both $\Rapt$ and $\v2$.

\section{Impact of the Equation of State}
As a last point, we explored the impact of the equation of state (EoS) on the 
observables and especially the correlation between $\Ra(q)/\Ra(g)$ and 
$v_2(p_T)$. This has to be taken as a first explorative study.

We notice that even if the free gas approximation is a quite reasonable approximation to describe
the relation between density and temperature for most of the evolution of the expanding QGP
it is not true close to $T_c$ where the relation between temperature and density is strongly modified by the cross-over phase transition. One generally thinks that for the description of
the high $p_T$ particles this can be discarded in first approximation. 
However once it is opened the way to the possibility of a quenching that is not 
proportional to the density but is stronger
close to the phase transition we will show that one should more carefully look into 
the problem.
In fact if quenching would be dominated by the $T\sim T_c$ region the time spent 
in this region by the fireball can be strongly modified if a realistic EoS is considered.
In fact for a simple 1+1D expansion
the relation between temperature and density is modified from Eq.(\ref{temp-time}) to:

\begin{equation}
\frac{T}{T_0}=\left(\frac{\rho}{\rho_0}\right)^{\beta(T)}=
\left(\frac{\tau_0}{\tau}\right)^{\beta(T)}
\label{rho-t-lQCD}
\end{equation}
where $\beta(T)$ is a temperature dependent coefficient that can be obtained from a fit to lattice QCD
data \cite{lQCD-EoS}. We have found:
\begin{equation}
	\beta(T)=\frac{1}{3}-a\left(\frac{T_c}{T}\right)^n \,\,\, \,\,, \,\,\,\,\, T\ge T_c
\end{equation}

with $a=0.15$ and $n=1.89$ and of course for $T>> T_c$ one gets $\beta \sim 1/3$.
The parameters has been calculated from a fit to the lattice QCD data of Ref.\cite{lQCD-EoS}
on the energy density and pressure.
We see that close to $T_c$ the $\beta$ coefficient is quite small which means that even if the density
goes down as $\tau^{-1}$, the temperature stays nearly constant 
as one can expect in a first order or strong cross-over phase transition. This means
that the system spends more time close to $T_c$ respect to the simple picture given by
Eq.(\ref{temp-time}) usually assumed. 

In order to estimate the impact of this correction we have performed a simulation for the 
$\Delta E_{loss}(T)$ behavior represented by the thick solid line in Fig.\ref{fig-dedtau}
which is similar to the delayed energy loss proposed by Pantuev \cite{Pantuev:2005jt} 
as a solution for the observed large elliptic flow. 
We consider only this case because it is of course the one that is much more affected
by the modification implied by Eq.(\ref{rho-t-lQCD}).
Instead for the opposite case of an $E_{loss}$ dominated by the high $T$ 
the effect is vanishing. We have again taken care to regulate the total energy
loss in order to have the correct amount of $\Rapt$.

The results are given by open symbols in Fig.\ref{v2_RAA_WS}. The open square is the result
without the inclusion of the flavor conversion mechanism and shows a further increase
of both $\v2$ and $\Ra(q)/\Ra(g)$, in line with the previously seen correlation. 
The enhancement of the $\v2$ in itself is due to the fact that with a EoS the system 
spend more time close to $T_c$, therefore in order to have the same amount of
suppression most of the energy loss occurs later.
However more interestingly
the inclusion of the $q \leftrightarrow g$ conversion results to be more efficient that in
the previous case generating a strong decrease of $\Ra(q)/\Ra(g)$ while keeping the same $\v2$.
The stronger effect of flavor conversion is again due to the longer lifetime.
The final result is a combination that move the final value of both $\v2$ and $\Ra(q)/\Ra(g)$
much close to the experimental ones shown by the shaded area in Fig.\ref{v2_RAA_WS}.
We also notice that a $\Ra(q)/\Ra(g) < 1$ is obtained while generally it is believed that flavor conversion by
inelastic collisions can give at most $\Ra(q)/\Ra(g) = 1$ \cite{Sickles_QM09}.
 
This last result about the impact of a realistic EoS deserves a more careful 
treatment again employing a more realistic description of the bulk.
Indeed a full description of the dynamics related to the cross-over region should
include also the gradual change from a quark-gluon plasma to an hadronic gas.
However this would be out of our aim for the present work.

\section{Conclusions}
In the present work we have considered different temperature
dependences of the energy loss tuning always the parameters to
reproduce the experimentally $\Rapt$ supressions of pions.
We have shown that even if the $\Rapt$ is fixed, different $\Delta E(T)$ 
generate very different value for both the $\v2$ and the $\Ra(q)/\Ra(g)$. 
Indeed both quantities are quite puzzling because standard jet quenching
models are not able to reproduce none of them. We refered to this 
as the "azimuthal" and "flavour" puzzle.
We have found however that there is a correlation between these two observables 
that is detemined by the temperature (or density) dependence of the quenching.
In agreement with Ref.\cite{LS_PRL,Pantuev:2005jt}
we have found that if the quenching is dominant closer to $T_c$ the $\v2$ is enhanced getting closer to the data.
However while in Ref.\cite{LS_PRL,Pantuev:2005jt} it is discarded a separate treatment of quark and gluons,
their $E_{loss}$ leads to a quite large ratio $\Ra(q)/\Ra(g)$
which would result incompatible with the observed systematics of 
$\Ra(p+\bar p)>\Ra(\pi^++\pi^-) \sim \Ra(\rho^0)$.
It appears that while the $\v2$ would suggest a 
$\Delta E(T)$ that increases towards $T_c$ the $\Ra(q)/\Ra(g)$ would become too large
even if we do not yet have a direct measurement of $\Ra(q)/\Ra(g)$ and we impinge
on the uncertainties coming from the fragmentation functions.

In this context we have also spot an unexpected strong dependence on
the density profile of the fireball that emerge for extreme choices of 
$\Delta E(T)$, while the
dependence is milder but not negligible for more standard energy loss like the GLV one.
This puts a warning for further studies: once going beyond the simple 
$\Rapt$ 
factor one has to rely on a realistic dynamical evolution of the bulk matter
like those supplied by hydrodynamics and/or parton cascades \cite{Fochler:2010wn}.

It seems that the only way to get closer to the observed values
for $\Rapt$, $\v2$ and $\Ra(q)/\Ra(g)$ solving both the "azimuthal" and "flavour" puzzles,
is to have both $\Delta E(T)$ increasing close to $T_c$ and a flavour conversion mechanism. 
Finally we point out that while generally the free gas approximation is quite reasonable to describe the
expansion of the QGP fireball, this is no longer true
if the energy loss occurs dominantly close to $T_c$. In such a case it has a significant impact to take into
account the strong deviation from the free gas approximation occuring in the cross-over region
and leading to a slowing down of the cooling close to $T_c$.
This increase further the time spent around $T_c$ which enhance both $\v2$ and the efficiency
of the $q\leftrightarrow g$ in medium conversion moving the values of $(\Ra(q)/\Ra(g),\v2)$ close to the
observed ones.

It is of course important to study how the longer lifetime and higher temperatures reached
at LHC energies affects the observed correlations.
Furthermore in Ref.\cite{Fries:2009ez} it has been pointed out that a better probe
of the flavor conversion mechanism is supplied by an high $p_T$ strangeness 
enhancement. It remains to be studied if such an enhancement is affected by the $T$
dependence of the energy loss.

In Summary we have pointed out the impact of peculiar $T$ dependences of the $E_{loss}$
on both $\v2$ and $\Ra(q)/\Ra(g)$ and their correlation. Moreover we have spot the relevance
that the $EoS$ may have in case of $E_{loss}$ dominated by the $T\sim T_c$ region.
In any case our study, although already revealing several interesting indications,
is mainly explorative. A more quantitative analysis should be performed
with more sophisticated models that include the energy loss fluctuations,
realistic gain and loss processes, elastic energy loss and a more accurate description of the bulk.

\end{document}